# Interpretations of family size distributions:
# The Datura example

Tomáš Henych*

Keith A. Holsapple

Aeronautics and Astronautics

University of Washington 352400

Seattle, WA 98195

*Corresponding author: ftom@physics.muni.cz

**Abstract**

Young asteroid families are unique sources of information about fragmentation physics and the structure of their parent bodies, since their physical properties have not changed much since their birth. Families have different properties such as age, size, taxonomy, collision severity and others, and understanding the effect of those properties on our observations of the size-frequency distribution (SFD) of family fragments can give us important insights into the hypervelocity collision processes at scales we cannot achieve in our laboratories.

Here we take as an example the very young Datura family, with a small 8-km parent body, and compare its size distribution to other families, with both large and small parent bodies, and created by both catastrophic and cratering formation events. We conclude that most likely explanation for the shallower size distribution compared to larger families is a more pronounced observational bias because of its small size. Its size distribution is perfectly normal when its parent body size is taken into account. We also discuss some other possibilities. In addition, we study another common feature: an offset or "bump" in the distribution occurring for a few of the larger elements. We hypothesize that it can be explained by a newly described regime of cratering, "spall cratering", which controls the majority of impact craters on the surface of small asteroids like Datura.

# 1. Introduction

Asteroid families are groups of asteroids of a common origin, first recognized by Hirayama (1918). They originate from material cast off from one minor body during an impact from another. That impact can be a cratering event, with a small part of the parent body being removed; or a catastrophic event, with a complete breakup into smaller pieces. Although the most prominent reported families were created from asteroids larger than 100 km, the discovery of families with smaller parent bodies has become possible using larger telescopes dedicated to discovering asteroids, by using more sophisticated numerical methods of searching for families in large datasets, and generally from a more concentrated effort of asteroid family research. That is very important because asteroid families are unique natural laboratories of breakup physics on scales we cannot study in our experimental facilities. The fragment properties contain information on the interior of the parent asteroid and provide us with important constraints of the asteroid physical properties. However, interpretations are required to extract that information. Those interpretations must be based on an understanding of how the family-forming event depends on the parameters of the bodies and the impact.

Our observations give us the size-frequency distribution (SFD) of family fragments. That distribution commonly has a number of recognizable characteristics – the separation of the second largest and the largest fragment in size, the slope of a first power-law part, a bend-over of the steep part to a shallower slope, an offset in sizes observed at larger fragments in some families and so on. All of these are probably an imprint of the break-up process and the properties of the parent asteroid. This information can be seriously confused by the presence of interlopers in the SFD. Those are asteroids with similar proper elements but no relation to a family whatsoever (Migliorini et al., 1995). There are several methods to remove interlopers from the family list which use various physical properties (albedos, colors, spectra) together with the proper elements to identify asteroid families without interlopers (Parker et al., 2008; Masiero et al., 2013; Carruba et al., 2013). Moreover, size dependence of the fragment ejection velocity and the Yarkovsky drift (Vokrouhlický et al., 2015) can put additional constraints on the family membership (Nesvorný et al., 2003).

Understanding those data will enable us to understand collisional physics at large scale, for various impact conditions, and for various asteroid types. This can help us create better fragmentation models for Solar System evolution models and, in turn, leads to a better description of the Solar System history. Here we consider and interpret the SFD of one young and interesting family with small members, the Datura family (Vokrouhlický et al., 2017).

# 2. The Datura family

The Datura family in the Main Belt (MB) was discovered and recognized as a very young family by Nesvorný et al. (2006). At that time, seven members of that family were known, 1270 Datura being the largest of them. Its age was calculated by numerically tracking family members' present trajectories backward in time resulting in a 450 ± 50 kyr estimate. That value was revisited by Vokrouhlický et al. (2009), and the current estimate is 530 ± 20 kyr. This relatively young age makes this family a very interesting group that can provide important insights into large-scale fragmentation and other physical processes that affect asteroids after their formation.

Rosaev & Plávalová (2015) found three new members of the Datura family, and later Vokrouhlický et al. (2017) presented a wealth of new data on that family: they extended its population to 17 members and two candidate members. They were able to derive spins for the six largest members and shapes of the four largest known members. The shape of the largest, 1270 Datura, was derived by Vokrouhlický et al. (2009). It is spinning rather fast (3.36-h period) and its shape is moderately elongated. On the other hand, the next three smaller members are all very elongated and slow rotators. Because the total mass of the known fragments compared to 1270 Datura is small, Vokrouhlický et al. (2017) concluded that the Datura family was formed by a cratering event in which the impact energy was much less than that for a catastrophic collision. They also derived a debiased fragment size distribution using a well-studied detection efficiency function of the Catalina Sky Survey. They noted that the size distribution of the Datura family is shallower than size distributions of other cratering families which originate from larger parent bodies.

*2.1 The Datura family size distribution*

Datura's size distribution is compared to other asteroid families in Fig. 1, along with the families in Vokrouhlický et al. (2017). We added one more cratering family, Aeolia, to present a wider range of family characteristics. We also show the debiased distribution as derived by Vokrouhlický et al. (2017). Note that we have chosen the distribution shown in the upper panel of their Fig. 16 to illustrate what the real distribution might look like. From the second largest and smaller, it is a two-part power-law distribution. The first part (larger fragment sizes) is steeper (the slope of −5.36 in this case) and the second part (smaller fragments) is shallower (the slope of −2.11). The transition (or break-point) size where the slope of the distribution changes is 1.08 km.

It is the flattening of the second part that we first consider. We recognize that some flattening of slope must be present in any size distributions at some small member size, because of the necessity of mass (volume) conservation. That is, the slopes of the family size cumulative distributions are usually quite steep, but a slope more negative than −3 cannot apply for increasingly smaller sizes, because the total volume of the fragments would then integrate to infinity. Thus there must be eventually some transition size at which the slope changes to some smaller value.

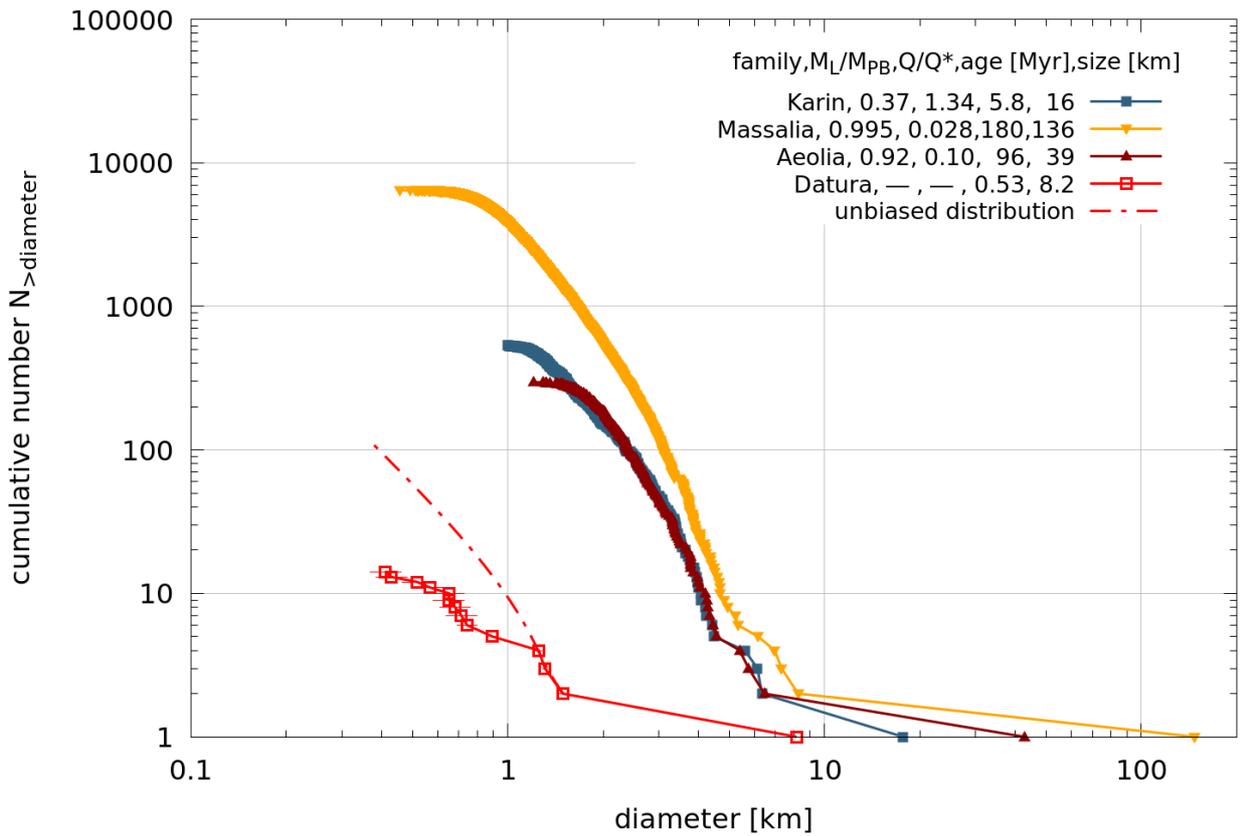

**Figure 1.** Observed (red open squares) and an example debiased (red dot-dashed line, Vokrouhlický et al., 2017) size distribution of the Datura family compared to larger-member families created by cratering, and to the Karin family produced by a catastrophic disruption. Note that Datura family members are much smaller than the majority of members of other families in the plot. In the plot legend we give estimates of the family's fraction of largest to parent body mass, severity of the impact as measured by the *Q/Q\** ratio, age and size of the largest member.

However, in this plot using the members' absolute sizes, it is apparent that other families also have a significant slope change beginning at about the same body size of about 2 km, not where the total family fragments mass approaches that of the putative parent body. That suggests that the feature is not real, but is simply due to an observational completeness limit due to natural biases of astronomical observations. Those biases are caused by the finite limiting magnitude of objects observable with our telescopes, resulting in a rapid decrease of the detection efficiency below that size. The fact that there remains some range of sizes at which the kick-over occurs for various families can easily be attributed to different albedos and distances at which those asteroids orbit the Sun in the Main Belt, both of which affect their brightness. So we propose that the shallower slope of the second part of the Datura family size distribution is not so surprising, because that shallow slope is the same as we observe in other families. It is due to the observational biases.

*2.2 Further comparisons*

A major factor affecting the SFD's is the severity of the impact. In terms of the impactor mass $m_p$, its velocity $U$ and the target mass $M$, the energy per unit mass of an impact is given by $Q=½m_pU^2/M$. That value can be compared to the specific energy $Q^*$ required to catastrophically disrupt and disperse the target body using the *qratio=Q/Q\**. The values of $Q^*$ are given in other studies, including in Holsapple et al. (2002). In Fig. 2 we give our current estimate of the $Q^*$ curve, an updated version of that in Holsapple et al. (2002) and compare it to some previous studies. There is great uncertainty in the "correct" value for the $Q^*$ curve, particularly in the gravity regime where there is no direct data. The $Q^*$ curve we chose is a theoretical curve justified by the laboratory experiments in strength regime (small-scale hypervelocity impacts), by the scaling theory that determines the slope in each regime and by the estimates of the specific energy needed for a breakup of three large asteroid families, Themis, Eos and Koronis (Fujiwara, 1982; Housen and Holsapple, 1990). Additionally, we used the largest impact craters observed on some asteroids of various sizes to give us the minimum specific energy of the breakup at that size to calibrate the $Q^*$ curve in gravity regime (Holsapple, 1994). We extended that data to include Šteins and Deimos; surprisingly Vesta is not so helpful in this respect. The equation for that curve is

$$Q^*_{J/kg} = \left(44\, r_{km}^{-0.6\mu} + 68\, r_{km}^{3\mu}\right)\left[\cos(\phi)\, U_{km/s}\right]^{2-3\mu} \qquad (1)$$

where $r$ is the asteroid radius in km, $U$ is the impact velocity in km/s, $\phi$ is an impact angle and $\mu$ is the point-source scaling-law exponent (e.g. Holsapple, 1993).

A small *qratio* indicates a cratering event with small mass removal, a value near unity is for an impact removing about 50% of the body, and a value much greater than unity indicates a severe pulverization. That *qratio* can be estimated by the mass difference between the largest member and the second and all smaller masses. For cratering events with a small *qratio*, the largest family member is very close to the size of the original parent body, and the second largest member and remaining elements are all much smaller than the largest one.

So, for further analyses, we compared Datura's size distribution to additional cratering families with a large parent body size range, for which we have some information on *qratio*.

We also limited ourselves to relatively young families with ages on the order of 100 Myr but not exceeding 500 Myr. We are not sure how important the time evolution of their size distribution is, but we have chosen quite large number of families and there are some very young families in our sample so we think our conclusions are robust. We address this problem for the Datura family in Sect. 3.2.

Table 1 gives the characteristics of asteroid families we have chosen for comparison with the Datura family. The main source of this family size distribution data is Nesvorný (2015), from which we removed some interlopers as noted in Nesvorný et al. (2015), Brož et al. (2013), Carruba & Nesvorný (2016) and Carruba et al. (2016).  In the table we give age in millions of years and the volume ratio of the largest family member with a diameter $D_L$ and the estimated volume of a parent body with a diameter $D_{PB}$ generally calculated from numerical simulations of the breakup. The *qratio* is then calculated from an assumed relation given as

$$qratio = Q/Q^* = \left[2 - 2\left(D_L/D_{PB}\right)^3\right]^{1+\mu/2} \qquad (2)$$

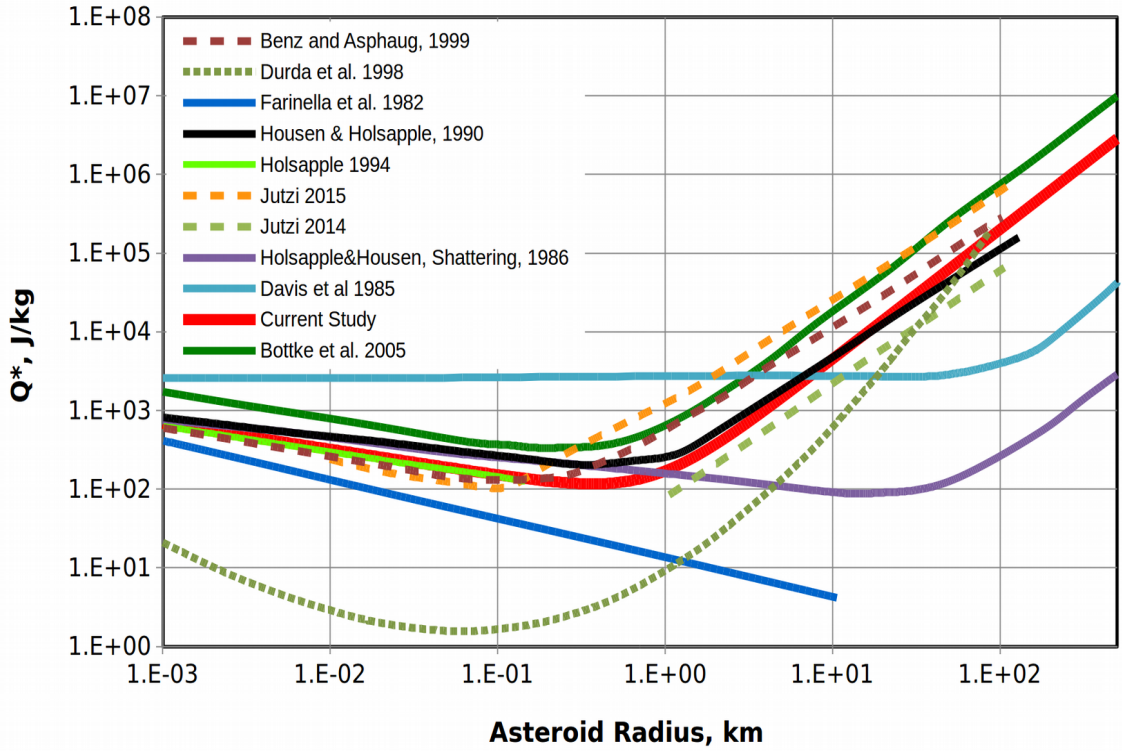

**Figure 2.** The $Q^*$ curve we use in the present study (thick red solid line), we used $\mu=0.55$ for non-porous rock-like materials in this plot. Shown in the plot are also estimates of the $Q^*$ from selected previous studies.

The value of $\mu$ used for a specific family is decided upon the spectral type of the largest member of that family (we assume, that the parent asteroid had the same spectral type): C and Ch types are assumed to be more porous which implies $\mu=0.4$ and for all other types (including ambiguous C/X) we used $\mu=0.55$ valid for non-porous rock-like or hard soil materials. The basis for this formula is the assumption that the largest mass is that of the parent body minus that of the crater. A similar relation was first derived by Housen et al. (1991) from the scaling laws and overpressure explosion laboratory data simulating asteroid breakup in gravity regime. This relation is also similar to an expression of Benz & Asphaug (1999) derived from their hydrocode simulations (see also Bottke et al., 2015).

In Table 1 we also give an average geometric albedo for the first 10–15 members and the diameter of the largest member $D_L$ where available, based on WISE/NEOWISE data (Mainzer et al., 2016). The information on families is taken from Nesvorný et al. (2003), Nesvorný et al. (2015), Brož et al. (2013), Spoto et al. (2015) and Carruba et al. (2016). We tried to remove interlopers based on the available information, but that is a complex problem which can only be solved by a synthesis of all available sources of information and that is a subject of some recent papers (Nesvorný et al., 2015 and references therein). All the size distributions we show in our plots should therefore be taken as tentative.

| family | age [Myr] | $(D_L/D_{PB})^3$ | $Q/Q^*$ | geometric albedo | spectral type | $D_L$ [km] |
|---|---|---|---|---|---|---|
| 2 Pallas | <500 | 0.966 | 0.032 | 0.167 | B | 514 |
| 3 Juno | 460 | 0.999 | 0.00036 | 0.155 | Sk | 247 |
| 20 Massalia | 180 | 0.995 | 0.0028 | 0.21 | S | 136 |
| 148 Gallia | <450 | 0.992 | 0.005 | 0.25 | S | 98 |
| 158 Koronis (2) | 10–15 | 0.987 | 0.01 | 0.24 | S | 39 |
| 163 Erigone | 212 | 0.68 | 0.59 | 0.061 | Ch | 72 |
| 298 Baptistina | 110 | 0.22 | 1.76 | 0.274 | C/X | 21 |
| 302 Clarissa | <100 | 0.96 | 0.04 | 0.06 | X | 39 |
| 363 Padua | 410 | 0.61 | 0.73 | 0.066 | Xc | 90 |
| 396 Aeolia | 96 | 0.92 | 0.1 | 0.097 | Xe | 39 |
| 434 Hungaria | 207 | 0.05 | 2.28 | 0.582 | E | 9 |
| 752 Sulamitis | <400 | 0.79 | 0.35 | 0.054 | C | 60 |
| 832 Karin | 5.8 | 0.37 | 1.34 | 0.19 | S | 16 |
| 945 Barcelona | <350 | 0.77 | 0.38 | 0.35 | S | 26 |
| 1128 Astrid | 135 | 0.52 | 0.95 | 0.062 | C | 42 |
| 1189 Terentia | <200 | 0.99 | 0.009 | 0.081 | Ch | 59 |
| 1222 Tina | <150 | 0.94 | 0.067 | 0.165 | X | 26 |
| 1270 Datura | 0.53 | — | — | 0.24 | Sk | 8.2 |
| 1521 Seinajoki | 160 | 0.19 | 1.84 | 0.173 | S | 20 |
| 1668 Hanna | 240 | — | — | 0.06 | Cg | 22 |
| 2344 Xizang | 220 | — | — | 0.153 | X/S | 17 |
| 2384 Schulhof | 0.78 | — | — | 0.28 | S | 11 |
| 7353 Kazuya | <100 | 0.32 | 1.47 | 0.191 | S | 11 |
| 10811 Lau | <100 | 0.34 | 1.43 | 0.274 | S | 8 |
| 14627 Emilkowalski | 0.22 | — | — | 0.201 | C/X | 7 |
| 16598 1992 YC2 | 0.15 | — | — | 0.153 | S | 5 |
| 18405 1993 FY12 | <200 | 0.22 | 1.77 | 0.155 | C/X | 9 |
| 18466 1995 SU37 | <300 | 0.09 | 2.14 | 0.254 | S | 6.4 |
| 21509 Lucascavin | 0.55 | — | — | 0.153 | S | 4 |

**Table 1.** Properties of asteroid families we used to compare with the Datura family. Age of the family in millions of years, volume ratio of the largest fragment and the parent body, *qratio* calculated by Eq. 2, geometric albedo used to calculate size of all the fragments from their absolute magnitude, spectral type and the size of the largest family member $D_L$ in km.

The size distribution of the Datura family compared to all of these other cratering families is shown in Fig. 3. We can see that the bend-over size to the flatter slope is again on the order of 1–2 km, but there is a somewhat larger dispersion, as might be expected from a wider range of albedos and distances. That is independent of the total mass of the fragments. The slope of the first part of the Datura size distribution is comparable to all these other cratering families.

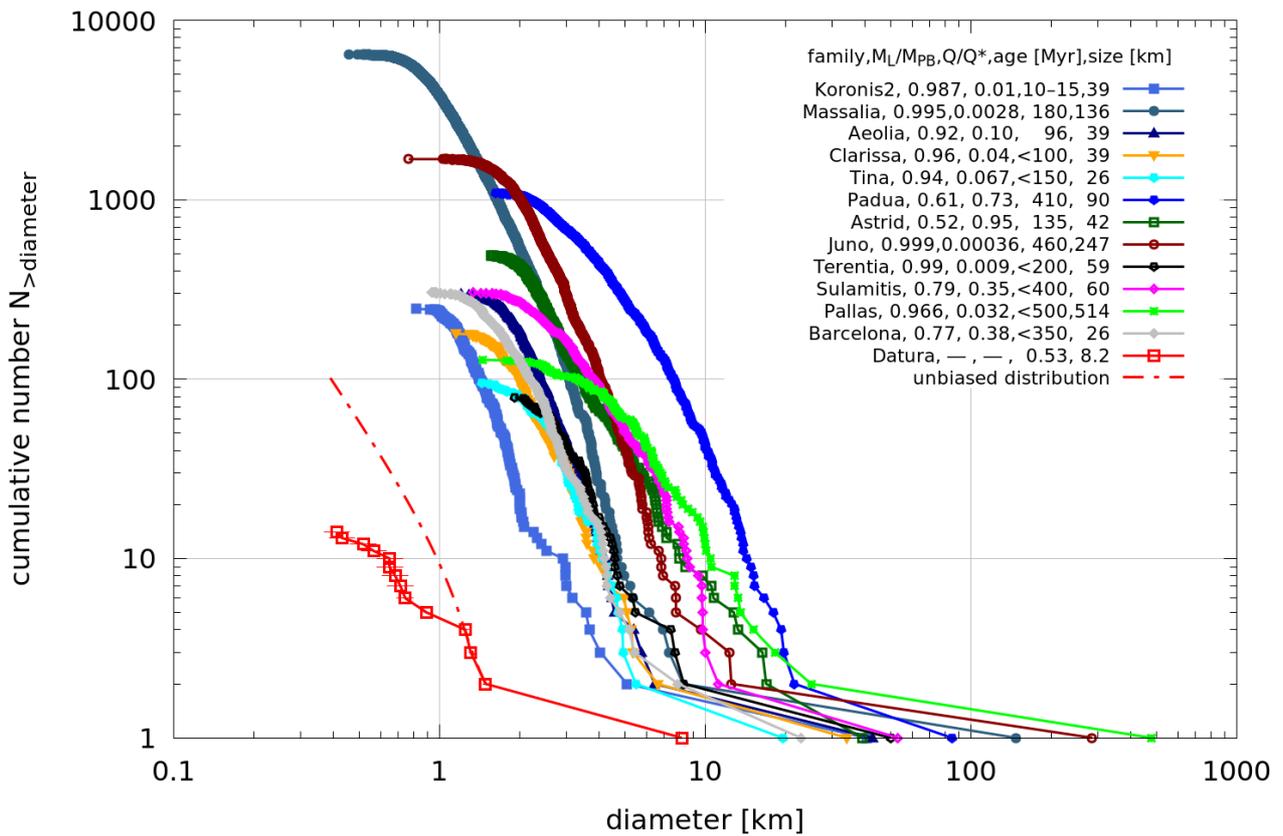

**Figure 3.** Size distribution of the Datura family compared to other cratering families. Compare both slopes of the debiased size distribution of the Datura family to others and also note where the transition to a shallower slope for all the families occurs.

Finally, in Fig. 4 we "zoom in" to a narrower range of parent body sizes and compare the Datura family size distribution to a much smaller span of member sizes, but now created by both cratering and catastrophic disruption impacts. All of those families are small in size, with the largest members being smaller than about 22 km. Some of them were discovered only recently and the number of their known members is very low, even lower than that of Datura (e.g., Lucascavin, 1992 YC2, Emilkowalski and Schulhof). Those we plot only because they are interesting targets for a future research since they seem to share some similarities with Datura and other such small-size families.

This comparison is also fairly consistent with our explanation of the shallow slope of the Datura's size distribution being due to the observational biases. The transition for those reasonably well-observed families again begins at the similar size of about 2 km. There is one peculiar distribution which is shallower than others, that of the Kazuya family. We note that it is a barely catastrophic family, but currently we have no explanation for the shallower slope of its size distribution.

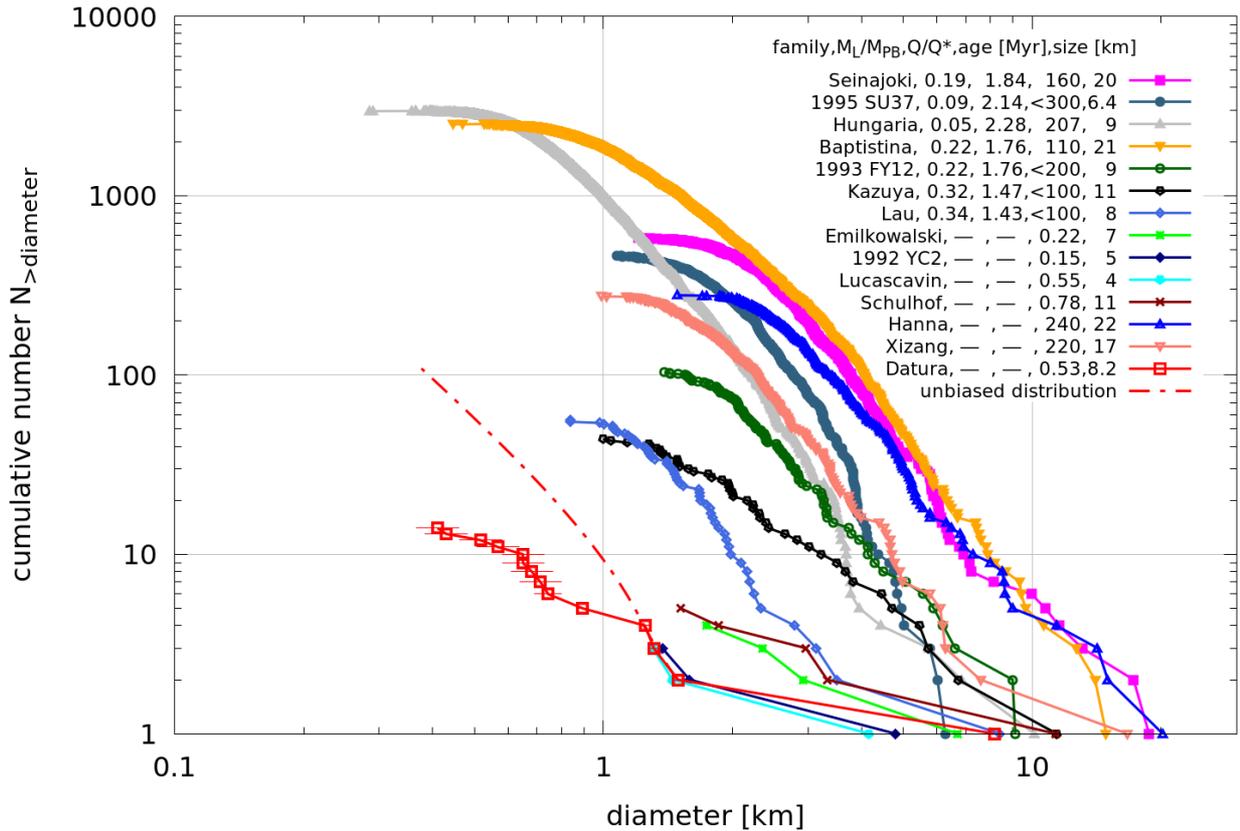

**Figure 4.** Size distribution of the Datura family compared to other small-parent-body families.

## 3. Other explanations

The explanation of the size distribution of the Datura family we give in Section 2 seems to be fairly robust, but we want to discuss other physical processes that have been suggested and that might be responsible for differences in the SFD.

*3.1 Cratering into a critically spinning asteroid*

Vokrouhlický et al. (2017) proposed that the different size distribution of the Datura family could be explained by an impact of a comparably smaller projectile onto a nearly critically-rotating parent body. Currently there are some laboratory experiments (Housen, 2004; Morris et al., 2012; Morris, 2015) that suggest the fast rotation could indeed substantially decrease the dispersion threshold $Q^*$ of an asteroid and that might affect the fragment size distribution. However, it is not possible to directly apply those lab experiments to Datura-sized asteroid when they happen in different impact regime than the laboratory experiments.

The current spin of the largest member has the period of 3.36 h. However, the original spin of the parent body could have been either faster or slower than that value. That original spin would be affected both by the location and angle of the impact and by the amount and speed of the material cast off. However, at the current spin, the reduction in gravity at the surface of an object rotating at a rate ω is proportional to $ω^2$ and the spin of Datura is only 64% of the critical gravity spin period limit of about 2.3 h. Thus, the reduction in gravity is only 40% or so. We doubt that such a spin would have a significant effect. However to repeat, we do not know the spin at the time of the family formation. Further study is required on that issue.

Further research in this area is certainly needed, presumably by the means of numerical simulations. Unfortunately, to date the reported SPH simulations of family formation do not have a fragmentation model capable of reproducing the initial fragment size distribution produced in laboratory experiments. Instead they always produce fragments equal to the size of their SPH particles of which there is commonly only about 40 in each 3D direction. Further developments in numerical modeling are needed.

*3.2 Collisional grinding of family members*

Any removal of a family element or a reduction in its size will flatten the observed SFD. One possibility is the removal of an element entirely by its catastrophic dispersion after the initial family formation. The importance of a catastrophic event can be estimated by a consideration of its lifetime.

The collisional lifetime of a 1-km asteroid is on the order of 0.5 Gyr, based on models of Bottke et al. (2005b), O'Brien & Greenberg (2005) and our own theoretical calculations. Therefore, in the 0.53 Myr Datura family age the probability of a family member being catastrophically dispersed is extremely low. For illustration, we give our average collisional lifetime estimates as a function of asteroid diameter together with some other estimates in Fig. 5.

But what about smaller family members? We calculated the probability of a catastrophic dispersion in 0.53 Myr using the present MB population for a diameter range of 0.01–10 km assuming both, strength and gravity controlled impact regimes, with a transition at 1 km. For simplicity, we assumed the asteroid size distribution was a two-part power law with a transition at 1 km. We used the incremental distribution from Bottke et al. (2005a, 2005b) to calculate the cumulative distribution and then we fitted a two-part power-law with a transition size at 1 km to that distribution. For smaller asteroids we obtained the slope of −2.6 while for larger bodies the slope was −2.04. The intrinsic probability of collision $P_i = 2.86 \cdot 10^{-18} \, \text{km}^{-2} \, \text{yr}^{-1}$ (Bottke et al., 1994; Bottke et al., 2005a), the impact velocity 5.3 km/s and the impact angle of 45°. For an asteroid of a given radius, we calculate the largest expected projectile from the Poisson statistics as

$$d_L = \left( P_i C_d r_{\text{km}}^2 t_{\text{yr}} \right)^{1/\alpha} \tag{3}$$

where $C_d = 1.2 \cdot 10^6$ is a constant from our idealized cumulative size distribution of Main Belt asteroids $N_{>d} = C_d d_{\text{km}}^{-\alpha}$ and $\alpha$ is the slope of this distribution, either for large or small asteroids. Then we calculate the projectile size that would already cause a catastrophic

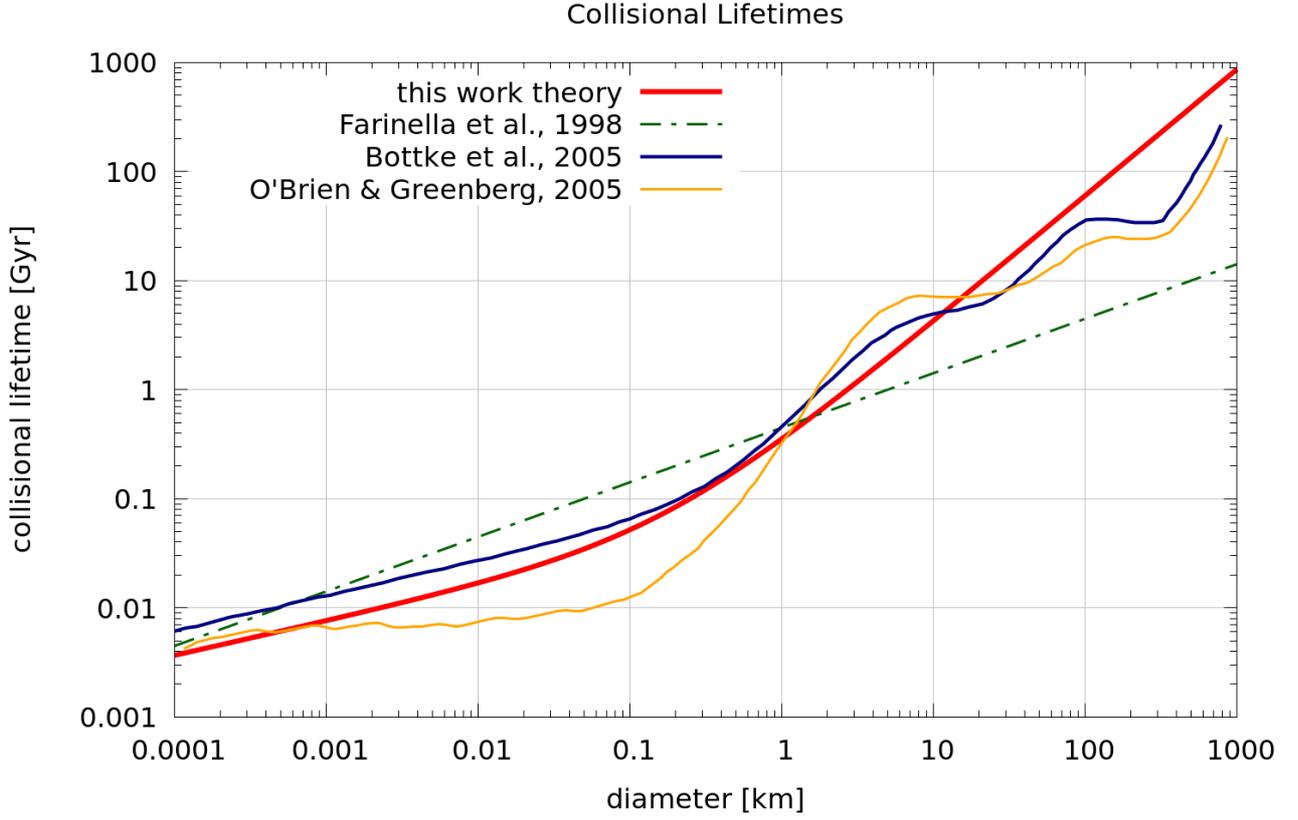

**Figure 5.** Average collisional lifetimes of MB asteroids. Our theoretical value based on two-slope power-law size distribution of asteroids is compared to previous numerical and theoretical models.

dispersion in either strength

$$\frac{d_{str}^*}{2 r_{km}} = 0.044 \, r_{km}^{-0.2\mu} U_{km/s}^{-\mu} \tag{4}$$

or gravity regime

$$\frac{d_{gr}^*}{2 r_{km}} = 0.051 \, r_{km}^{\mu} U_{km/s}^{-\mu} \tag{5}$$

These were derived from the Eq. 1 that describes our $Q^*$ curve. From the two diameters we can calculate the probability of getting disrupted using Poisson statistics

$$P = 1 - \exp\left(-\left(\frac{d^*}{d_L}\right)^{-\alpha}\right) \tag{6}$$

where we use the appropriate $d^*$ in strength or gravity regime and also the appropriate slope of the population $\alpha$. When we set $d_L = d^*$ and solve for time, we can calculate the average collisional lifetime

$$t_{coll} = \left(0.082 \, r_{km}^{0.314} + 0.64 \, r_{km}^{1.16}\right) \text{Gyr} \tag{7}$$

Based on these values the resulting probability of a catastrophic breakup in 0.53 Myr for a 0.01-km asteroid was about 7%, for a 0.3-km asteroid it was about 2% and for a 1-km it was 0.2%. Therefore, in the short time after the Datura family formation, only about one out of fifty of its small 0.3-km family members could be expected to be disrupted by an impact. We conclude that it is very unlikely that a collisional disruption affected the Datura family size distribution observed to date.

*3.3 Rotational fission of family members*

There is another effect that might lead to a breakup of small asteroids on rather short timescales. A gradually increasing spin leading to a rotational fission can be caused by the Yarkovsky–O'Keefe–Radzievskii–Paddack (YORP) effect. This can create a binary asteroid or an asteroid pair (Margot et al., 2015) and could potentially affect the size distribution of family members (Jacobson et al., 2014). We calculated the timescale of the YORP spinup for the Datura family members smaller than 1 km based on the measured spin-up of 6 small MB asteroids, usually explained as being the result of the YORP effect, and scaled the result for the average heliocentric distance of the Datura family members (Vokrouhlický et al., 2015). Although we did not know the thermal properties of those asteroids, we assumed they be the same as for the asteroids with a detected spinup (half of them has the S spectral type, the same as we assume for all Datura family members). From this calculation we estimated the timescale of the spinup from 8-h spin period to 2-h spin period be on the order of several tens of Myr and therefore two orders of magnitude longer than the estimated age of the Datura family. We can therefore conclude that the Datura family size distribution was not affected by the YORP spin-up.

## 4. Bumps and additional features

There are other interesting features observed in the Datura and other family SFD's that provide clues about their formation. There is in many cases a characteristic offset or "bump" in the SFD of the Fig. 1 families, often between about the fourth and the smaller fragments, but always for only a small number of members. That would seem to indicate a transition in fragment formation mechanics. One is present in the Datura family SFD for the second through $5^{th}$ fragments. But in most families those occur at a fixed member number, not its size. Therefore it is not an observational bias.

We can think of some possible explanations for that, including differences in reaccumulation, fragment spins, or one or more missing members caused by a body's disruption. We also understand that some of those large family members could be interlopers, but it is very unlikely that this would be the case for all of those families. And none of those ideas would seem to account for the fact that in many families it occurs at about the $4^{th}$ largest member. That feature also appears in other families as shown below. We plot several cratering and catastrophic dispersion families that have this feature in their size distributions in Fig. 6 and 7, respectively. We can see that it appears for wide range of collision severity, largest member size and for various family ages.

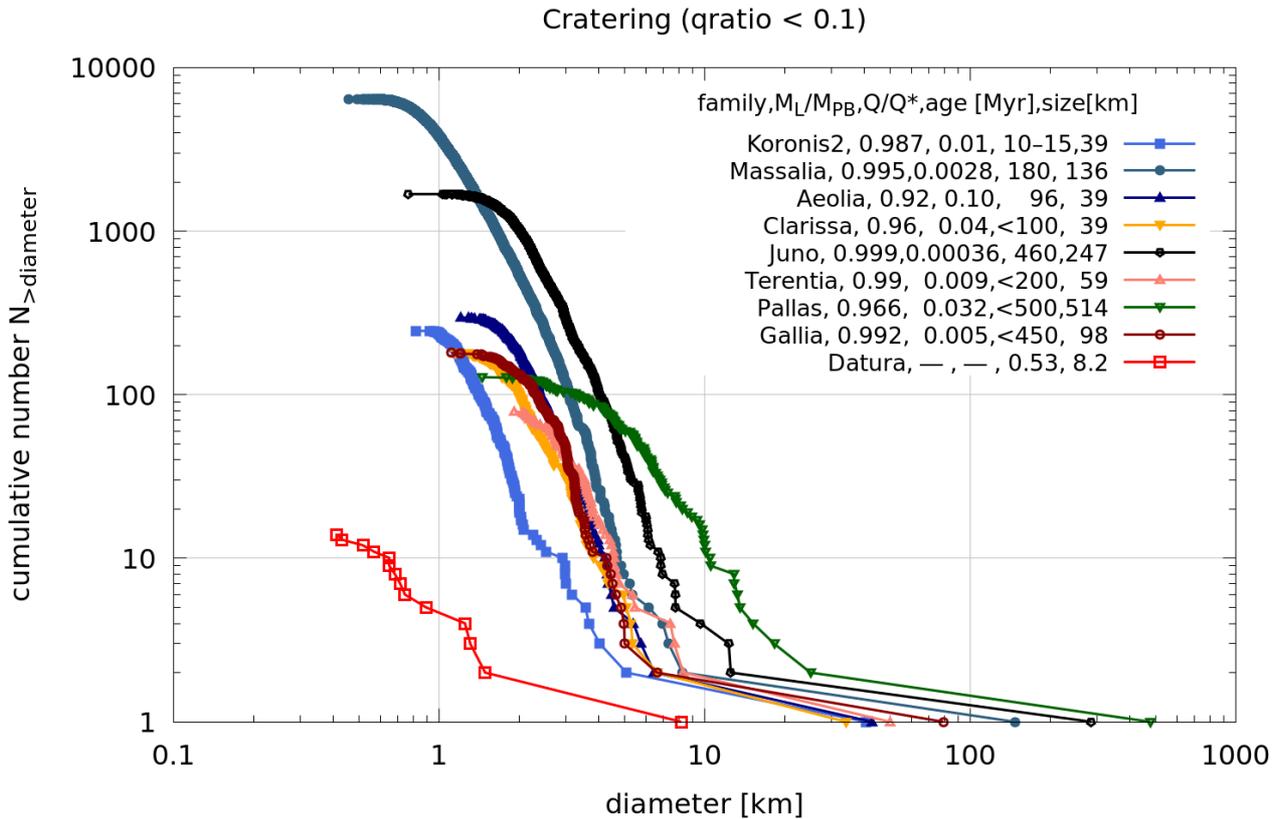

**Figure 6.** Some cratering families with a "bump" in their size distributions near the fourth fragment.

It could partly be caused by the way we determine the family size distributions, namely that we use the average albedo for all family members. This can potentially cause some calculated size discrepancies and it could be especially problematic for more heterogeneous families. To test this, we also plotted absolute magnitude distributions for those families and the feature was still present, although for some families it was slightly less pronounced.

More likely these bumps are a real feature that has to do with the impact process and it could be an important constraint on the breakup models. If it is real for the Datura family, then the debiased distribution could actually look different, namely the steep power-law part could be shifted to the left and start from the sixth member of that family. We now introduce a new possible explanation for those bumps.

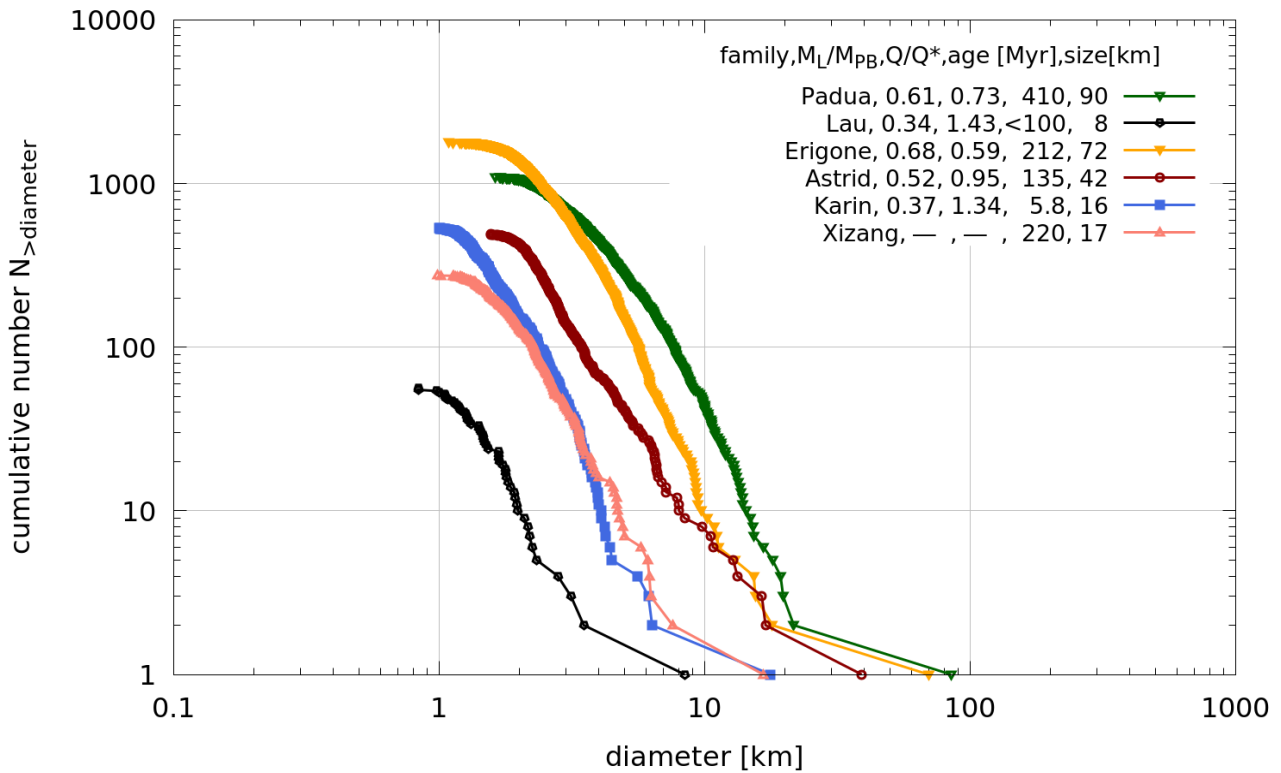

**Figure 7.** Large cratering and catastrophic breakup families with a "bump" in their size distributions near the fourth fragment.

*4.1 Spall cratering*

Clearly the SFD of a family depends strongly on the mechanics of impacts. There is the well-known and studied demarcation between strength-dominated and gravity-dominated events. But there is also an additional important subset of the strength regime for smaller events called the "spall regime" that has been recently identified[1] (Holsapple & Housen, 2013).

In the spall regime a crater is formed primarily by spallation (tensile fracture parallel to the surface) of near-surface plates of material. The result is a very shallow wide spall crater with a small central pit crater formed by the more usual shearing flows. Fig. 8 illustrates a spall crater with the spall plates inserted back into the crater.

Those spall craters are of minor consequence for all larger terrestrial and lunar impacts, but they are the result of *all* laboratory experiments in rocky brittle material and most likely for all craters on small rocky asteroids. Not only the crater, but especially the fragments of those experiments are very different from those of the classical bowl-shaped craters that we observe on the Moon, Mercury, Earth and other large bodies.

---

1A detailed manuscript of the mechanics and scaling in the spall regime is in preparation by Holsapple and Housen.

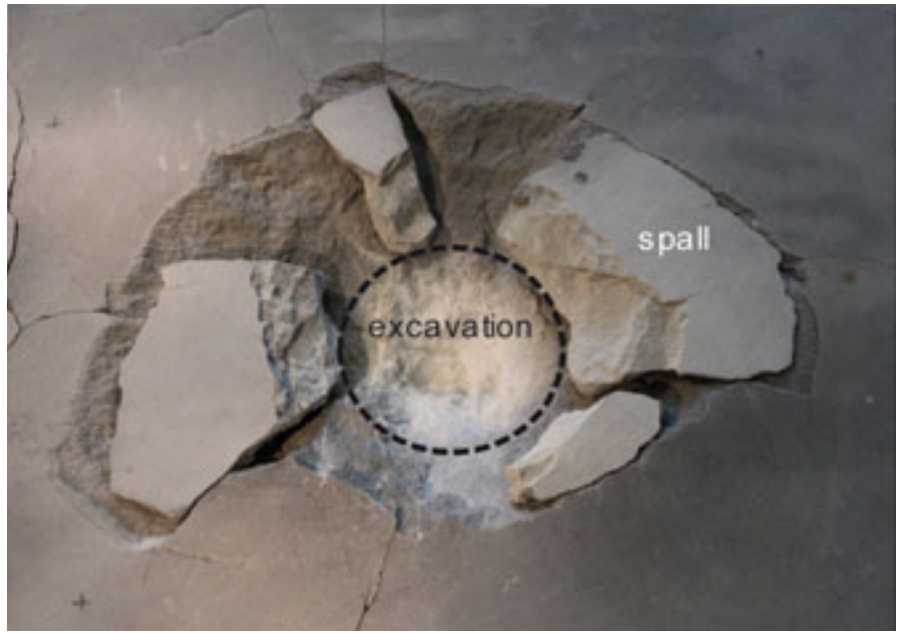

**Figure 8.** Typical spall crater from a laboratory experiment (Dufresne et al., 2013). There are a few large fragments that are much larger than any from the formation of the central bowl excavation crater.

The initial fragment size distribution from spall craters has some relatively small number of much larger fragments than the large number of small fragments from a "regular" crater. In addition, those larger fragments might have more reaccumulation than smaller fragments. Those fragments could explain the peculiar offset of a few larger family elements discussed above. We were intrigued by that idea, and spend some time studying that possibility.

Holsapple & Housen (2013) presented a plot showing the range of surface gravity (body diameter) and crater size where spall craters could be expected. An updated estimate is given here as Fig. 9.

Since Datura is a small 8-km asteroid, with a surface gravity of 0.003 m/s$^2$, almost every impact that does not disrupt it would produce spall craters. The estimated severity of the Datura impact (small *qratio*) is well below that limit. So, the Datura family initial SFD would be expected to have a few relatively large fragments. In addition, the largest fragments are flat and elongated bodies, all very suggestive of spall fragments. Therefore, we hypothesize that the Datura family was produced by spall cratering, and that is the cause for the bump after the first three elements. Note however, that offsets are also evident for families that were created by much more severe impacts, so those might be due to other reasons. Further study is needed.

To test this hypothesis for Datura, we compared the Datura size distribution to size distributions of cratering laboratory experiments that were known spall events. We do so, assuming that impact crater on Datura family parent body was also a spall crater. In Fig. 10 we plot fragment size distributions of several laboratory experiments together with the size distribution of the Datura family.

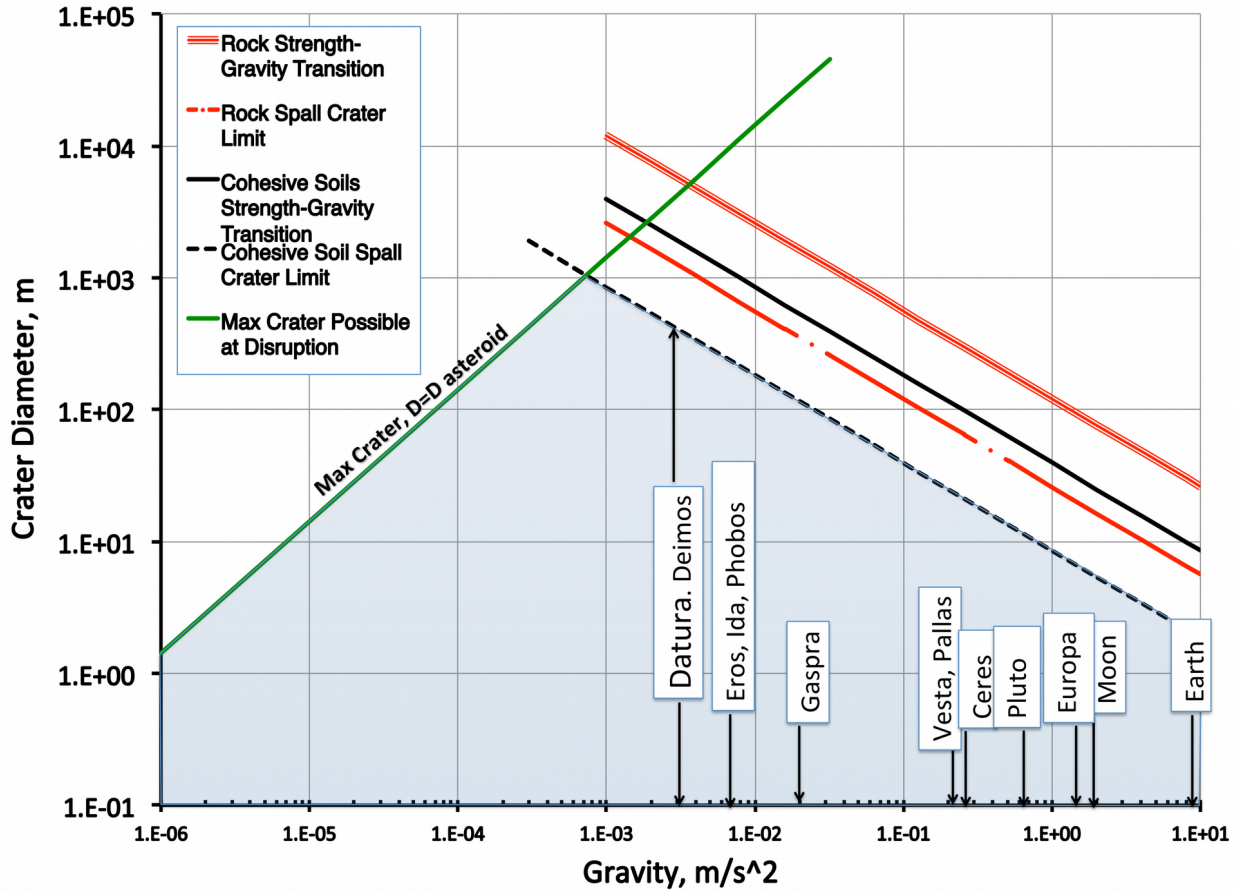

**Figure 9.** Estimates of spall cratering, strength cratering, and gravity cratering regimes as determined by the surface gravity (which is set by the object diameter) and the crater size. The green upward sloping line at the left defines the largest crater possible without disrupting the object. All possible craters on a body of on the order of 10–20 km will be spall craters.

    All of the lab distributions exhibit the bump with 3–7 elements. The fragment size distribution of the laboratory experiment s2570 of Michikami et al. (2016), which was a spall event, is very similar to Datura's distribution, and we think that the *qratio* of the collision that produced the Datura family is only slightly smaller than the *qratio* of 0.061 of that experiment.

    Unfortunately, the comparison is marred because most of those experiments used cubic or rectilinear targets instead of smoothly rounded ones, and that can especially affect fragment size distributions. Reflection of shock waves preferentially cuts the sharp corners and forms large fragments.

    Further support for our hypothesis comes from the shapes of the few largest fragments (as derived from inversion of their lightcurves) which are very elongated. That is also the same as the spall fragments that we observe in laboratory experiments. While some of the most elongated fragments may later be identified as contact binary asteroids, those could simply be from two spall fragments that escaped the family on a very similar trajectory and with similar velocity, or, alternatively, from two broken spall fragments that ended up as a close or a contact binary.

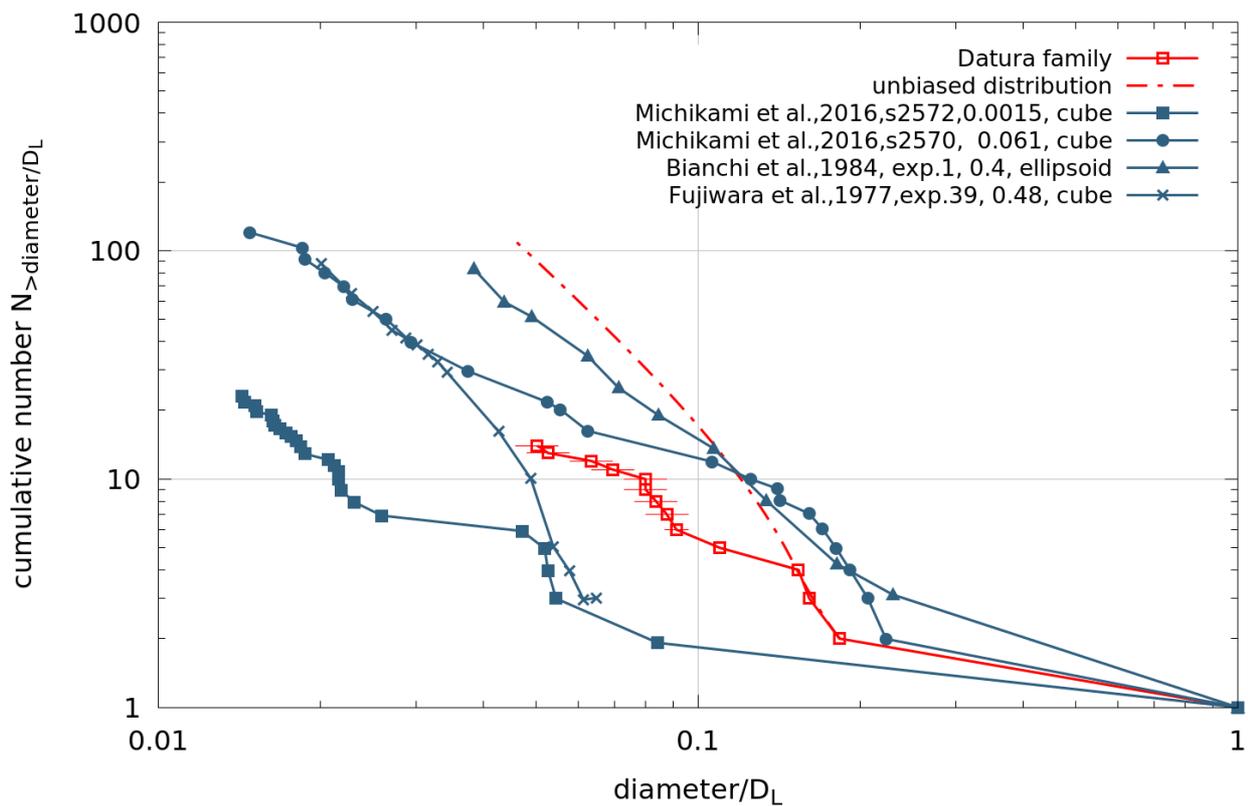

**Figure 10.** Fragment size distribution of the Datura family compared to fragment size distributions of cratering laboratory experiments. For those we give the reference and the name of the experiment, its *qratio*, and the shape of the sample used in that experiment.

## 5. Conclusions

We are studying impact events in the Main Belt by observing the details of the family size-frequency distributions. We use the Datura family as our example and compare it to other families. We conclude that the relatively low slope of its SFD is simply due to observational bias, which is a consequence of the relatively small size of the largest fragment of that family. Members of the Datura family (except for Datura itself) are as small as the smallest members of other families we currently observe. The observation of the asteroids at the size of about 1–2 km is heavily biased and therefore even for other families we do not see their real size distribution.

There are several other families that are comparable in size to the Datura family and it is most desirable that we find new members of these families in future asteroid datasets from the planned or already commissioned telescopes that focus on asteroids (Pan-STARRS, Gaia or LSST). Their size distributions and other properties could bring new light into fragmentation physics of such small asteroids and that could lead to improved fragmentation models in the evolution models studying history of the Solar System.

Spall cratering is newly identified interesting sub regime of strength dominated impacts, it is manifested by a different morphology of craters and fragment size distributions. We hypothesize that the offset in fragment sizes observed in some families including the Datura family may be caused by spall cratering. From the theory we expect that for an 8-km asteroid majority of the craters on its surface would be spall craters. That would explain the elongated shapes of the largest fragments that we observe in that family. Also the Datura family size distribution is similar to laboratory cratering experiments that were known spall events. Since spallation is caused by tensile failure, direct observation or an indirect evidence for spall cratering on asteroids can give us important constraints on tensile strength of asteroid outer layers.

## Acknowledgments

This work was supported by NASA grant NNX12AG18G. Comments from two anonymous referees are greatly appreciated as they helped us to substantially improve the manuscript.